\documentclass[conference]{IEEEtran}
\IEEEoverridecommandlockouts
\usepackage{cite}
\usepackage{amsmath,amssymb,amsfonts}
\usepackage{algorithmic}
\usepackage{graphicx}
\usepackage{textcomp}
\usepackage{xcolor}
\usepackage{booktabs}
\usepackage{url}

\def\BibTeX{{\rm B\kern-.05em{\sc i\kern-.025em b}\kern-.08em
    T\kern-.1667em\lower.7ex\hbox{E}\kern-.125emX}}
\begin{document}
\title{RiverEcho: Real-Time Interactive Digital System for Ancient Yellow River Culture}

\author{
\small
\textit{Haofeng Wang}$^{*}$, \textit{Yilin Guo}$^{\dagger}$, \textit{Zehao Li}$^{\ddagger}$, \textit{Tong Yue}$^{\dagger}$, 
\textit{Yizong Wang}$^{\dagger}$,
\textit{Enci Zhang}$^{*}$, \textit{Rongqun Lin}$^{\S}$,
\textit{Feng Gao}$^{\P}$\thanks{Feng Gao and Siwei Ma are Corresponding authors}, \textit{Shiqi Wang}$^{**}$, \textit{Siwei Ma}$^{\dagger}$ \\
$^{*}$\textit{School of Electronic and Computer Engineering, Peking University}, Shenzhen, China \\
$^{\dagger}$\textit{School of Computer Science, Peking University}, Beijing, China \\
$^{\ddagger}$\textit{The School of History, Renmin University of China}, Beijing, China \\
$^{\S}$\textit{Pengcheng Laboratory}, Shenzhen, China \\
$^{\P}$\textit{School of Arts, Peking University}, Beijing, China \\
$^{**}$\textit{Department of Computer Science, City University of Hong Kong}, Hong Kong SAR, China
}
\maketitle

\begin{abstract}
The Yellow River is China's mother river and a cradle of human civilization. The ancient Yellow River culture is, moreover, an indispensable part of human art history. To conserve and inherit the ancient Yellow River culture, we designed RiverEcho, a real-time interactive system that responds to voice queries using a large language model and a cultural knowledge dataset, delivering explanations through a talking-head digital human. Specifically, we built a knowledge database focused on the ancient Yellow River culture, including the collection of historical texts and the processing pipeline. Experimental results demonstrate that leveraging Retrieval-Augmented Generation (RAG) on the proposed dataset enhances the response quality of the Large Language Model(LLM), enabling the system to generate more professional and informative responses. Our work not only diversifies the means of promoting Yellow River culture but also provides users with deeper cultural insights. 
\end{abstract}


\section{Introduction}
\label{sec:intro}
Historical documents from the Yellow River Basin record the millennia-long development of Chinese civilization, laying a solid foundation for the economic and cultural progress of later generations\cite{CaoGuangzhang2022HistoricalInheritanceAndContemporaryValueOfYellowRiverCulture}. However, the dissemination of Yellow River culture among the general public remains limited. One of the primary challenges lies in the inherent difficulty of collecting and digitizing ancient materials, which are often fragmented, widely dispersed, and preserved in non-digital formats. Moreover, the complexity of historical carriers and the difficulty in interpreting ancient texts further hinder public access and understanding. Therefore, it is essential to leverage rapidly advancing technologies to overcome these barriers and enable the broad dissemination of this cultural heritage.

Nowadays, human-computer interaction (HCI) systems have been widely applied across fields such as healthcare, education, and law \cite{langote2024human, de2021teaching, amato2022intelligent}, benefiting the public and significantly enhancing efficiency and user experience in various industries. However, research and applications concerning the preservation and transmission of traditional culture remain relatively scarce \cite{hirsch2024human}, with efforts to preserve the ancient Yellow River culture being virtually nonexistent. As a vital symbol of Chinese history and identity, protecting the ancient Yellow River culture urgently requires greater attention, and the development of HCI systems may offer innovative support for its preservation and inheritance.

This work presents RiverEcho, a real-time interactive intelligent system for ancient Yellow River culture, which integrates Automatic Speech Recognition (ASR), Large Language Models (LLM), Text-to-Speech (TTS), and Talking-head generation. The system is capable of recognizing users’ spoken questions, generating culturally and historically informed responses, and dynamically presenting these answers through a talking-head virtual human. The entire system employs a streaming pipeline architecture, enabling low-latency responses and ensuring real-time interaction and response coherence.

To enable the system to accurately address historical and cultural inquiries raised by users and to effectively disseminate the historical and cultural knowledge related to the ancient Yellow River region, while also mitigating the risk of hallucinated content from the LLM, this work constructs a dedicated high-quality knowledge dataset for ancient Yellow River culture. Specifically, we collected over 100 ancient manuscripts from different Chinese dynasties and disciplinary domains related to the Yellow River, along with authoritative works authored by leading contemporary experts in Yellow River cultural studies. These materials were subjected to automated preprocessing and hierarchical annotation, followed by manual sampling verification and supplementary labeling conducted by history students. As a result, we obtained a curated text corpus containing over 20,000 fully annotated cultural and historical segments. Based on this corpus, we incorporated a Retrieval-Augmented Generation (RAG) mechanism into the LLM’s inference pipeline, significantly enhancing the factual accuracy and cultural relevance of the generated responses.

\begin{figure*}[htbp]
    \vspace{-12pt}
    \centering
    \includegraphics[width=0.9\textwidth]{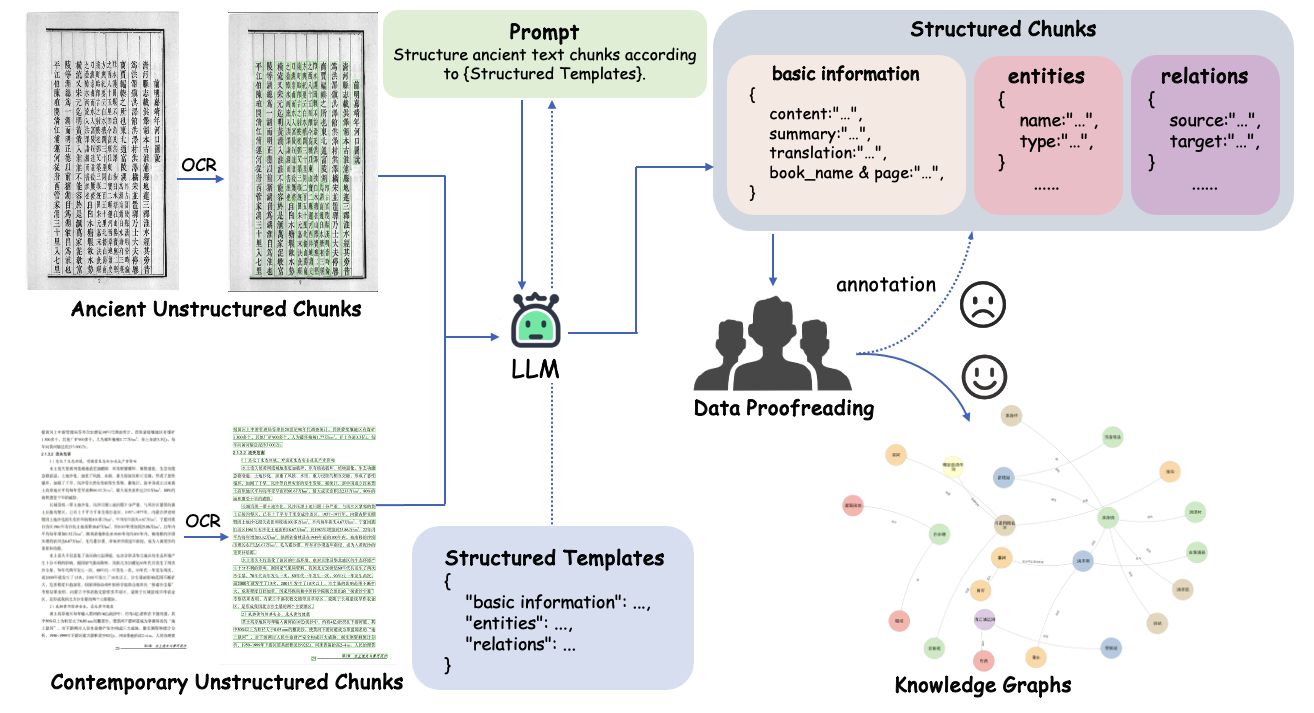}
    \caption{\textbf{Overall process of dataset construction.} First, the collected ancient and contemporary books are processed to obtain unstructured chunks, which are then fed into an LLM along with prompts containing structured templates to generate structured chunks. Next, the structured chunks undergo manual data proofreading. Chunks that do not meet quality standards are annotated and sent back for re-proofreading. Finally, the verified structured chunks are transformed into knowledge graphs and stored in the system.}
    \label{dataset}
\end{figure*}

Meanwhile, considering Li Daoyuan's remarkable contributions to ancient waterway systems, we used his historical persona as an example to construct a digital human-driven system. This enhances the overall visual appeal of the framework while also stimulating user engagement and meeting their visual expectations. 

Overall, our contributions are summarized as follows:
\begin{itemize}
\item[$\bullet$]We built a knowledge database focused on the ancient Yellow River culture, including the collection of historical texts and the processing pipeline, which has promoted the digitization of Yellow River-related documents and laid the foundation for in-depth cultural research and exploration.

\item[$\bullet$]We developed RiverEcho, a digital system capable of real-time interaction with users, which is driven by digital human images and can provide real-time and accurate answers to user-inputted questions, especially those related to the Yellow River.

\item[$\bullet$]We execute a thorough series of experiments to validate the effectiveness of the proposed dataset. The results demonstrate that leveraging RAG in our dataset enhances the response quality of large language models, enabling the system to generate more professional and informative responses.
\end{itemize}

\section{Related works}
\subsection{Large Language Model}
Nowadays, Large Language Models (LLMs)\cite{achiam2023gpt, touvron2023llama, yang2023baichuan, liu2024deepseek, yang2024qwen2} are emerging rapidly. They have demonstrated 
remarkable capabilities across a wide range of natural language processing (NLP) tasks. Through instruction tuning and RAG, contemporary LLMs have achieved not only impressive general intelligence but also notable expertise in specialized domains\cite{roziere2023code, li2023chatdoctor, zhang2024baichuan4, cui2023chatlaw}. This has spurred growing interest in the development of domain-specific LLMs, a rapidly evolving area of research that continues to drive innovation and exploration within the field.

The rapid development of LLMs has contributed to some extent to the dissemination and preservation of Chinese traditional culture. Recent research has focused on Classical Chinese Understanding (CCU), aiming to enhance the comprehension and generation of classical texts. Early CCU systems were primarily designed for specific tasks, such as translation\cite{jiang2023towards, chang2021time}, punctuation restoration\cite{li2009punctuation}, and named entity recognition (NER) \cite{yu2020bert,han2018cnn}. Recent 
\begin{figure*}[htbp]
    \vspace{-12pt}
    \centering
    \includegraphics[width=0.9\textwidth]{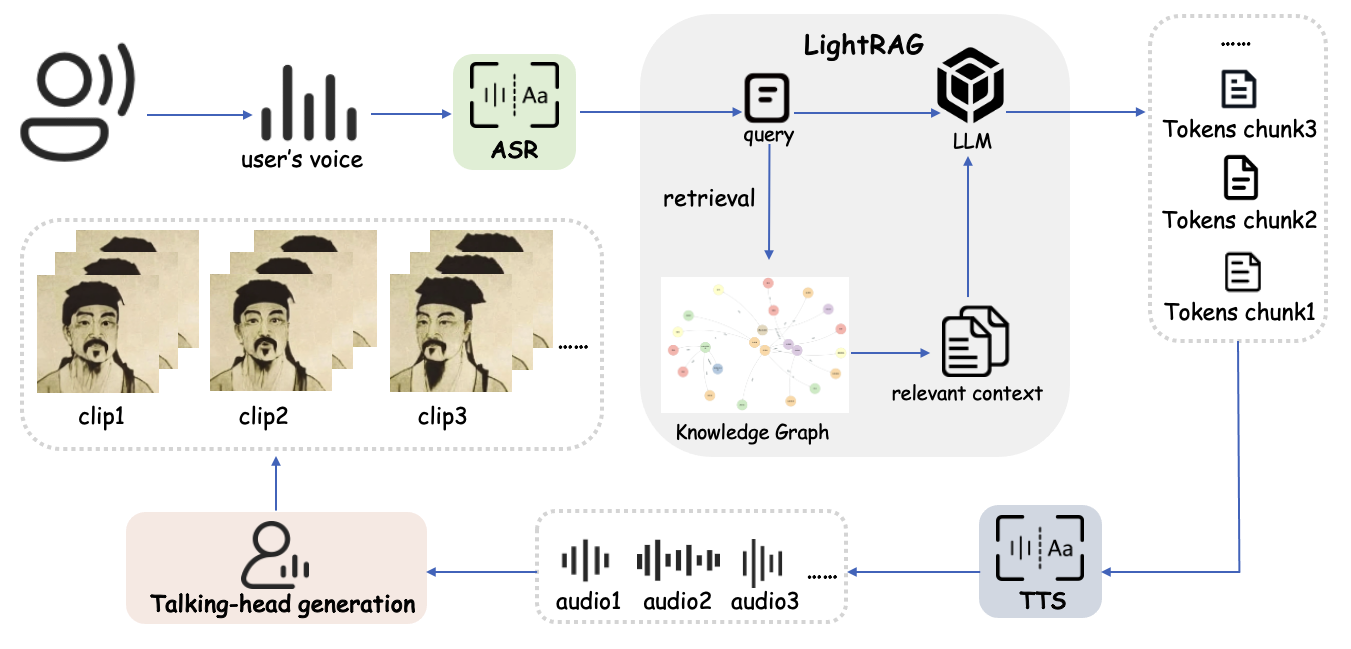}
    \caption{\textbf{Overall architecture of RiverEcho.} The user's voice queries is first converted into a textual query by the ASR module. Using LightRAG, relevant context is retrieved from the constructed knowledge graph and combined with the query as input to the LLM, which then generates token chunks in a streaming manner. The TTS module receives these token chunks and synthesizes corresponding audio segments in real-time, which are then fed into the talking-head generation module to produce video clips for presentation.}
    \label{overall_arc}
\end{figure*}
advancements, such as GujiBERT\cite{wang2023gujibert}, have utilized large-scale unlabeled classical Chinese corpora for masked pre-training, providing task-specific models with embeddings enriched with classical Chinese knowledge. Similarly, SikuGPT \cite{chang2023sikugpt} has demonstrated the potential of generative pre-training for classical poetry and prose creation. Additionally, models like Bloom-7b-Chunhua \cite{wptoux2023bloom} and Xunzi-Qwen-7B-CHAT \cite{xunzillm2024} have combined open-source base models with extensive classical Chinese corpora, offering preliminary insights into the capabilities of LLMs in understanding and generating classical Chinese texts. Notably, TongGu\cite{cao2024tonggu}, in the same work, proposed a two-stage fine-tuning approach, enabling the model to perform multiple tasks such as classical text reading comprehension, classical-to-modern Chinese translation, and classical poetry generation. These developments highlight the growing potential of LLMs in the preservation and inheritance of cultural heritage.

Nevertheless, integrating ancient Yellow River culture with LLMs is far from straightforward. The main challenge is the absence of a structured dataset specifically built around Yellow River classics and scholarly texts, making it difficult for LLMs to accurately capture and interpret the rich historical content embedded in this tradition. This gap underscores the necessity of developing dedicated datasets and tailored methods to facilitate a deeper understanding of this vital component of Chinese cultural heritage.
\subsection{Real-time Interactive Digital Humans for Audio-Visual Dialogue}
Recent advancements in real-time interactive digital human technology have enabled synchronous audio and video dialogue. LiveTalking\cite{livetalking} introduces a framework for audio-video synchronization, supporting models like wav2lip\cite{wav2lip} and musetalk\cite{zhang2024musetalk} to address initial inference delays. Metahuman-stream\cite{metahumanstream} offers an open-source solution with voice cloning, speech interruption, and full-body video stitching, compatible with RTMP\cite{rtmp} and WebRTC\cite{webrtc} protocols. Synthesia\cite{synthesia} provides a platform for creating digital avatars with lifelike narration in over 140 languages, reducing video production time significantly. VTube Studio\cite{vtubestudio} enables real-time avatar control using face tracking, enhancing interactive experiences through open-source integration. These open-source technical frameworks lay a solid foundation for the development of domain-specific applications.

\section{Methodology}
To enable the system to provide more professional and information-dense responses related to Yellow River culture, we constructed an Ancient Yellow River Cultural Dataset. In addition, we implemented a real-time interactive digital system, in which we adopted a RAG framework to enhance the response quality of the LLM using the constructed dataset.

\subsection{Ancient Yellow River Cultural Dataset}

\textbf{Ancient Book Collection:}
The historical documents related to the Ancient Yellow River culture have been preserved in various forms, such as manuscripts, bamboo slips, and inscriptions, with their textual content recorded in Classical Chinese, Tangut script, and other writing systems. The diversity of these media, along with the challenges in interpreting historical scripts, has created a significant barrier to public understanding of the Ancient Yellow River culture.

To facilitate the widespread dissemination of this cultural heritage, we have collected a diverse set of historical texts based on recommendations from scholars specializing in Chinese history. The proposed dataset includes ancient texts from various historical periods (Pre-Qin, Han, Wei-Jin and Northern and Southern Dynasties, Tang-Song, and Ming-Qing periods), primarily consisting of modern annotated editions. Additionally, it encompasses over a hundred historical and contemporary works related to the Yellow River, covering multiple themes: river governance, technology and engineering, natural knowledge, socio-economic aspects, cultural heritage, historical narratives, disasters and their impacts, and interdisciplinary topics.
\begin{table}[!ht]
    \centering
    \caption{Distribution of Themes in Proposed Dataset}  
    \begin{tabular}{c c}  
        \toprule
        \textbf{Theme} & \textbf{Number of Trunks} \\  
        \midrule
        \textbf{Total} & \textbf{20408} \\ 
        River governance & 6125 \\
        Technology and engineering & 4369 \\
        Natural knowledge & 2552 \\
        Socio-economic aspects & 1649  \\
        Cultural heritage & 1778 \\
        Historical narratives & 1551 \\
        Disasters and their impacts & 1268 \\
        Interdisciplinary topics &  1116 \\ 
        \bottomrule
    \end{tabular}
    \label{tab:example}
\end{table}

\textbf{Data Structuring and Processing Pipeline:}
As shown in Figure ~\ref{dataset}, the construction of the dataset consists of three steps. First, the collected large-scale ancient books are preprocessed using vertical text OCR to obtain unstructured text. Next, an LLM is utilized to structure the unstructured text. Finally, the structured data undergoes a proofreading process to ensure accuracy. We elaborate on the last two steps in detail.

\textit{Unstructured Data Structuring:}
To improve the retrieval efficiency and enhancement capability of the LLM when processing the dataset, we performed structured processing on the collected Yellow River historical texts. We first divided the collected unstructured text into multiple unstructured chunks. Then, using a large LLM along with a corresponding structuring template, we converted these unstructured chunks into structured data, performed entity deduplication, and constructed a knowledge graph. The structured chunks consist of three key components: 
\begin{itemize}
    \item \textbf{Basic information}: Includes the original text, translation, and summary, as well as the corresponding book title and page number.
    \item \textbf{Entities}: Refer to the named entities and their types that appear within the paragraph.
    \item \textbf{Relations}: Represent the connections between different entities.
\end{itemize}

\textit{Data Proofreading:}
To enhance the professionalism of responses to Yellow River-related inquiries, we incorporated a human proofreading mechanism. Specifically, we invited professional reviewers to conduct randomized sampling and verification of the structured data. They were requested to identify, assess, and annotate instances of hallucination, including incorrect translations, overgeneralizations, and excessive information supplementation.
To enhance the reliability and consistency of the verification process, each flagged hallucination case was subjected to a two-stage human review pipeline. In the first stage, a student reviewer annotated the detected errors and categorized them based on predefined error types. In the second stage, another reviewer reassessed the annotations, verified the discrepancies, and finalized the manual corrections.

\subsection{Real-Time Interactive Digital System}
As illustrated in Figure~\ref{overall_arc}, our real-time interactive system integrates multiple key modules, including ASR, LLM, TTS, and talking-head generation.

Specifically, the system first utilizes the ASR module to convert a user's voice queries into text. Then, the system employs the LightRAG\cite{guo2024lightrag} technique to perform retrieval-based matching within the Ancient Yellow River Cultural Database. The retrieved chunks, along with the original query, are subsequently fed into the LLM as contextual inputs, enabling the model to generate a streaming text response. Next, the TTS module converts this streaming text into real-time speech, which is then synthesized into a talking-head video and transmitted to the front end for display.

Since the latency of existing ASR tools is generally negligible, we do not apply additional streaming processing to it. However, all other modules in the system operate in a streaming fashion, except for the LLM. Given that both the inputs (except for the LLM) and outputs of all modules are streamed, the system follows a fully integrated streaming workflow. In other words, once the ASR module transcribes the user's speech, all subsequent modules function in parallel. This design significantly reduces response latency, particularly when generating long sequences of tokens, ensuring a highly efficient real-time user interaction experience.

Specifically, we adopt Livetalking\cite{livetalking}, an open-source real-time interactive digital human platform, as the basic framework of our system. For each module, we adopt the following configurations:

\begin{itemize}
    \item \textit{ASR}: We use FunASR\cite{gao2023funasr}, a high-performance open-source ASR model, capable of recognizing user input speech with both high speed and accuracy.
    \item \textit{LLM}: We employ Qwen2.5-max\cite{yang2024qwen2} as the base model and integrate LightRAG\cite{guo2024lightrag} for database retrieval and enhanced inference, optimizing the system's performance in information retrieval tasks. The tokens generated by the LLM will accumulate into chunks until encounter punctuation marks that indicate the end of a sentence, such as a period, exclamation mark, or ellipsis.
    \item \textit{TTS}: We use Edge-TTS\cite{edge-tts}, an open-source, powerful few-shot speech synthesis model capable of generating high-quality speech with limited data. The sample rate is configured at 16,000 Hz, the voice is selected as "zh-CN-YunjianNeural," and the synthesis speed is adjusted to 20\% slower than the default.
    \item \textit{Talking-Head Generation Module}: We implemented the training and synthesis of ancient digital characters using the MuseTalk\cite{zhang2024musetalk} framework, an audio-driven lip-synchronization system designed for real-time high-fidelity facial animation. Specifically, we gathered dozens of portraits and statues of Li Daoyuan from the Internet and extracted their facial features. Leveraging generative models, we synthesized a series of modernized character images. We then refined these images by manually selecting and processing their micro-expressions to construct a compact character dataset. Finally, utilizing MuseTalk, we developed and animated digital humans based on this dataset.
\end{itemize}

Our system seamlessly integrates these modules to enable efficient streaming processing, delivering a smooth and responsive low-latency interactive experience for users. Optimizing the coordination between components, ensures real-time performance with minimal delays, enhancing overall usability.
\section{Experiments}
\subsection{Effectiveness of Proposed Dataset}
To demonstrate the enhancement effect on the professionalism and information richness of responses to Yellow River culture-related inquiries by LLMs through RAG using our proposed dataset, we selected Qwen2.5\cite{yang2024qwen2} as the baseline model and formulated ten Yellow River-related questions covering different topics. Additionally, we introduced two other well-established LLMs\cite{liu2024deepseek, achiam2023gpt} for a horizontal comparison.

\begin{figure*}[htbp]
    \centering
    \includegraphics[width=1\textwidth]{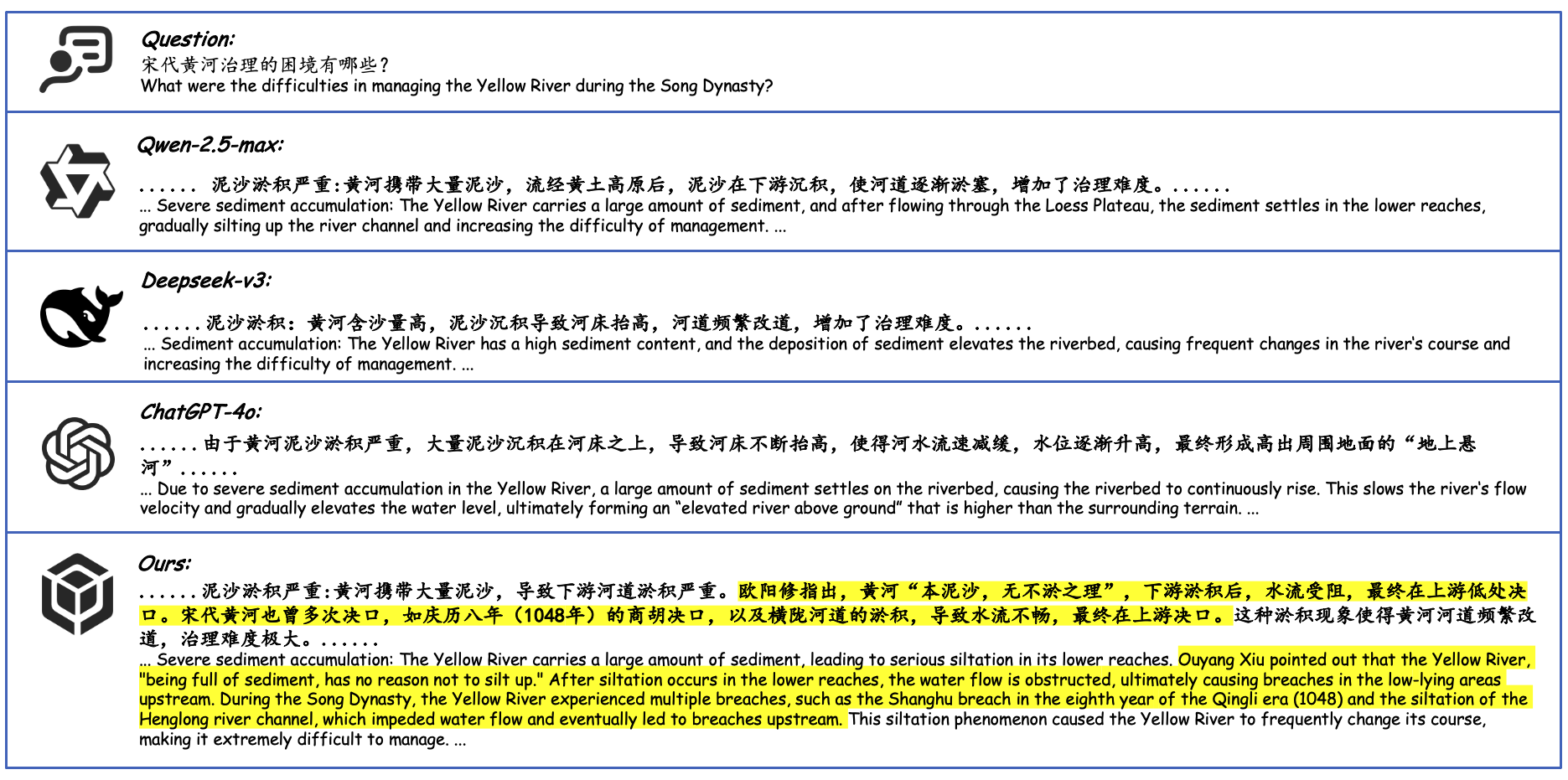}
    \caption{\textbf{An example comparing the responses of different models}, where the yellow-highlighted enhanced text is retrieved from the dataset via RAG. It can be seen that our dataset enhances the responses with a professional and rich context.}
    \label{comparison of models}
\end{figure*}
\begin{figure}[htbp]
    \centering
    \includegraphics[page=1, width=1.0\linewidth]{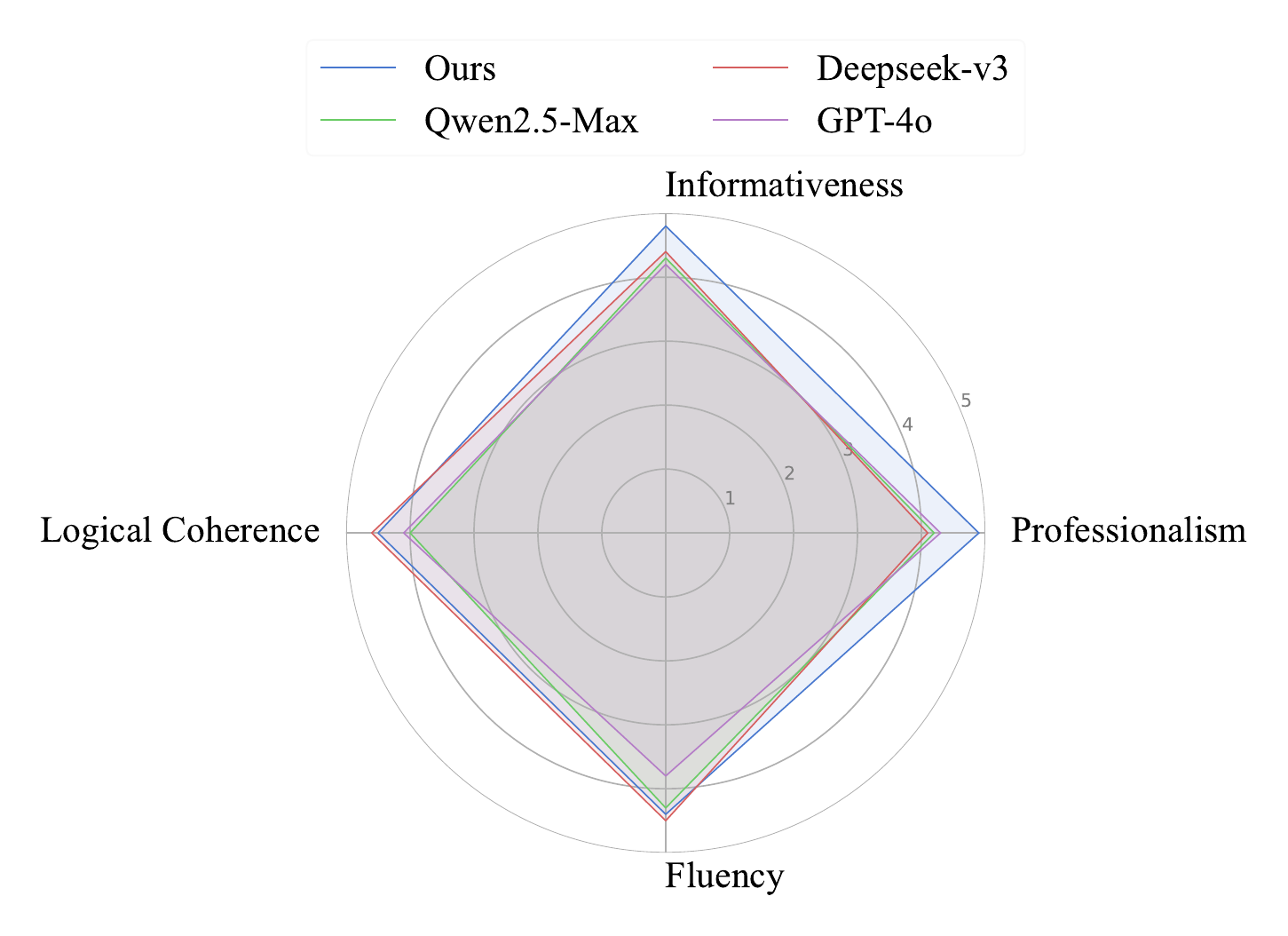}
    \caption{Subjective score comparison of different models.}
    \label{score_lidar}
\end{figure}

\begin{table*}[htbp]
    \centering
    \caption{Processing Time of Each Module in the System}  
    \begin{tabular}{c c c c}  
        \toprule
        \textbf{Module} & \textbf{Model} & \textbf{Processing Metric} & \textbf{Processing Time} \\  
        \midrule
        ASR & FunASR\cite{gao2023funasr} & Time required to recognize 1s audio & 0.01460 s \\ 
        LLM (including RAG) & Qwen2.5\cite{yang2024qwen2} + RAG & Tokens generated per second & 36.79 tokens/s \\
        TTS & Edge-TTS\cite{edge-tts} & Time required to synthesize 1s audio & 0.27448 s \\
        Talking-Head Generation & MuseTalk\cite{zhang2024musetalk} & Time required to drive one frame & 0.0039 s \\
        \bottomrule
    \end{tabular}
    \label{tab:processing_time}
\end{table*}
To evaluate the responses, we conducted a subjective experiment among over 100+  by assessing them from four key dimensions: professionalism, informativeness, logical coherence, and fluency. Specifically, \textit{fluency} refers to the speed of the output generated by the large language model. As shown in Figure~\ref{score_lidar}, our model achieved significantly higher scores in professionalism and informativeness compared to other models. Furthermore, as illustrated in Figure~\ref{comparison of models}, we selected a representative question to showcase the effectiveness of our model. Our model successfully leveraged retrieved texts (highlighted in yellow) to generate a more informative and enriched response.


\subsection{Response Time Analysis}
To validate the real-time performance of our system, we measured the latency of the entire system and its individual modules. Our system was run on an NVIDIA 4090 graphics card, occupying approximately 11.7GB of GPU memory. Table~\ref{tab:processing_time} presents the latency of each module in our system. Specifically, the processing latency of the ASR module is negligible; the LLM module, including the RAG process, can generate nearly 37 tokens per second; the TTS module requires only 0.27448 seconds to generate 1 second of audio; and the final Talking-Head Generation module achieves an audio-driven speed of 25 fps. Thanks to the streaming input and output of most modules, the system we proposed exhibits excellent real-time performance. 

\section{Conclusion}
In this paper, we present RiverEcho, a real-time interactive digital system designed for the Ancient Yellow River culture. It processes user voice queries and delivers professional, informative responses in real-time via a digital human interface. Specifically, to enhance the output performance of the LLM module, we collected and processed historical texts and modern books related to the Ancient Yellow River from different dynasties and various topics, constructing a dataset for Ancient Yellow River culture. Finally, we conducted a subjective evaluation to validate the effectiveness of this system. We hope that artificial intelligence will contribute to the preservation, revitalization, and innovative dissemination of traditional Chinese culture.

\section*{ACKNOWLEDGMENT} 
This work was supported by the National Key R\&D Program of China 2022YFF0902400, NSFC 62025101, BNSF L242014 and New Cornerstone Science Foundation through the XPLORER PRIZE.

\bibliographystyle{IEEEbib}
\bibliography{icme2025references}

\begin{thebibliography}{10}

\bibitem{CaoGuangzhang2022HistoricalInheritanceAndContemporaryValueOfYellowRiverCulture}
Guangzhang Cao,
\newblock ``The historical inheritance and contemporary value of yellow river culture,''
\newblock {\em Jinyang Academic Journal}, , no. 2, 2022.

\bibitem{langote2024human}
Meher Langote, Saniya Saratkar, Praveen Kumar, Prateek Verma, Chetan Puri, Swapnil Gundewar, and Palash Gourshettiwar,
\newblock ``Human-computer interaction in healthcare: Comprehensive review.,''
\newblock {\em AIMS Bioengineering}, vol. 11, no. 3, 2024.

\bibitem{de2021teaching}
Lizette De~Wet,
\newblock ``Teaching human-computer interaction modules—and then came covid-19,''
\newblock {\em Frontiers in Computer Science}, vol. 3, pp. 793466, 2021.

\bibitem{amato2022intelligent}
Flora Amato, Leonard Barolli, Giovanni Cozzolino, Antonino Ferraro, and Marco Giacalone,
\newblock ``An intelligent interface for human-computer interaction in legal domain,''
\newblock in {\em International Conference on P2P, Parallel, Grid, Cloud and Internet Computing}. Springer, 2022, pp. 240--248.

\bibitem{hirsch2024human}
Linda Hirsch, Siiri Paananen, Denise Lengyel, Jonna H{\"a}kkil{\"a}, Georgios Toubekis, Reem Talhouk, and Luke Hespanhol,
\newblock ``Human--computer interaction (hci) advances to re-contextualize cultural heritage toward multiperspectivity, inclusion, and sensemaking,''
\newblock {\em Applied Sciences}, vol. 14, no. 17, pp. 7652, 2024.

\bibitem{achiam2023gpt}
Josh Achiam, Steven Adler, Sandhini Agarwal, Lama Ahmad, Ilge Akkaya, Florencia~Leoni Aleman, Diogo Almeida, Janko Altenschmidt, Sam Altman, Shyamal Anadkat, et~al.,
\newblock ``Gpt-4 technical report,''
\newblock {\em arXiv preprint arXiv:2303.08774}, 2023.

\bibitem{touvron2023llama}
Hugo Touvron, Thibaut Lavril, Gautier Izacard, Xavier Martinet, Marie-Anne Lachaux, Timoth{\'e}e Lacroix, Baptiste Rozi{\`e}re, Naman Goyal, Eric Hambro, Faisal Azhar, et~al.,
\newblock ``Llama: Open and efficient foundation language models,''
\newblock {\em arXiv preprint arXiv:2302.13971}, 2023.

\bibitem{yang2023baichuan}
Aiyuan Yang, Bin Xiao, Bingning Wang, Borong Zhang, Ce~Bian, Chao Yin, Chenxu Lv, Da~Pan, Dian Wang, Dong Yan, et~al.,
\newblock ``Baichuan 2: Open large-scale language models,''
\newblock {\em arXiv preprint arXiv:2309.10305}, 2023.

\bibitem{liu2024deepseek}
Aixin Liu, Bei Feng, Bing Xue, Bingxuan Wang, Bochao Wu, Chengda Lu, Chenggang Zhao, Chengqi Deng, Chenyu Zhang, Chong Ruan, et~al.,
\newblock ``Deepseek-v3 technical report,''
\newblock {\em arXiv preprint arXiv:2412.19437}, 2024.

\bibitem{yang2024qwen2}
An~Yang, Baosong Yang, Beichen Zhang, Binyuan Hui, Bo~Zheng, Bowen Yu, Chengyuan Li, Dayiheng Liu, Fei Huang, Haoran Wei, et~al.,
\newblock ``Qwen2. 5 technical report,''
\newblock {\em arXiv preprint arXiv:2412.15115}, 2024.

\bibitem{roziere2023code}
Baptiste Roziere, Jonas Gehring, Fabian Gloeckle, Sten Sootla, Itai Gat, Xiaoqing~Ellen Tan, Yossi Adi, Jingyu Liu, Romain Sauvestre, Tal Remez, et~al.,
\newblock ``Code llama: Open foundation models for code,''
\newblock {\em arXiv preprint arXiv:2308.12950}, 2023.

\bibitem{li2023chatdoctor}
Yunxiang Li, Zihan Li, Kai Zhang, Ruilong Dan, Steve Jiang, and You Zhang,
\newblock ``Chatdoctor: A medical chat model fine-tuned on a large language model meta-ai (llama) using medical domain knowledge,''
\newblock {\em Cureus}, vol. 15, no. 6, 2023.

\bibitem{zhang2024baichuan4}
Hanyu Zhang, Boyu Qiu, Yuhao Feng, Shuqi Li, Qian Ma, Xiyuan Zhang, Qiang Ju, Dong Yan, and Jian Xie,
\newblock ``Baichuan4-finance technical report,''
\newblock {\em arXiv preprint arXiv:2412.15270}, 2024.

\bibitem{cui2023chatlaw}
Jiaxi Cui, Munan Ning, Zongjian Li, Bohua Chen, Yang Yan, Hao Li, Bin Ling, Yonghong Tian, and Li~Yuan,
\newblock ``Chatlaw: A multi-agent collaborative legal assistant with knowledge graph enhanced mixture-of-experts large language model,''
\newblock {\em arXiv preprint arXiv:2306.16092}, 2023.

\bibitem{jiang2023towards}
Zongyuan Jiang, Jiapeng Wang, Jiahuan Cao, Xue Gao, and Lianwen Jin,
\newblock ``Towards better translations from classical to modern chinese: A new dataset and a new method,''
\newblock in {\em CCF International Conference on Natural Language Processing and Chinese Computing}. Springer, 2023, pp. 387--399.

\bibitem{chang2021time}
Ernie Chang, Yow-Ting Shiue, Hui-Syuan Yeh, and Vera Demberg,
\newblock ``Time-aware ancient chinese text translation and inference,''
\newblock {\em arXiv preprint arXiv:2107.03179}, 2021.

\bibitem{li2009punctuation}
Zhongguo Li and Maosong Sun,
\newblock ``Punctuation as implicit annotations for chinese word segmentation,''
\newblock {\em Computational Linguistics}, vol. 35, no. 4, pp. 505--512, 2009.

\bibitem{yu2020bert}
Peng Yu and Xin Wang,
\newblock ``Bert-based named entity recognition in chinese twenty-four histories,''
\newblock in {\em International Conference on Web Information Systems and Applications}. Springer, 2020, pp. 289--301.

\bibitem{han2018cnn}
Xiaowei Han, Lizhen Xu, and Feng Qiao,
\newblock ``Cnn-bilstm-crf model for term extraction in chinese corpus,''
\newblock in {\em Web Information Systems and Applications: 15th International Conference, WISA 2018, Taiyuan, China, September 14--15, 2018, Proceedings 15}. Springer, 2018, pp. 267--274.

\bibitem{wang2023gujibert}
Dongbo Wang, Chang Liu, Zhixiao Zhao, Si~Shen, Liu Liu, Bin Li, Haotian Hu, Mengcheng Wu, Litao Lin, Xue Zhao, et~al.,
\newblock ``Gujibert and gujigpt: Construction of intelligent information processing foundation language models for ancient texts,''
\newblock {\em arXiv preprint arXiv:2307.05354}, 2023.

\bibitem{chang2023sikugpt}
Liu Chang, Wang Dongbo, Zhao Zhixiao, Hu~Die, Wu~Mengcheng, Lin Litao, Shen Si, Li~Bin, Liu Jiangfeng, Zhang Hai, et~al.,
\newblock ``Sikugpt: A generative pre-trained model for intelligent information processing of ancient texts from the perspective of digital humanities,''
\newblock {\em arXiv preprint arXiv:2304.07778}, 2023.

\bibitem{wptoux2023bloom}
Wptoux,
\newblock ``Bloom-7b-chunhua,'' \url{https://huggingface.co/wptoux/bloom-7b-chunhua}, 2023,
\newblock Accessed: 2023-10-01.

\bibitem{xunzillm2024}
Xunzi-LLM of~Chinese-classics,
\newblock ``Xunziallm,'' \url{https://github.com/Xunzi-LLM-of-Chinese-classics/XunziALLM}, 2024,
\newblock Accessed: 2024-03-01.

\bibitem{cao2024tonggu}
Jiahuan Cao, Dezhi Peng, Peirong Zhang, Yongxin Shi, Yang Liu, Kai Ding, and Lianwen Jin,
\newblock ``Tonggu: Mastering classical chinese understanding with knowledge-grounded large language models,''
\newblock {\em arXiv preprint arXiv:2407.03937}, 2024.

\bibitem{livetalking}
Lipku,
\newblock ``{LiveTalking}: Real-time interactive streaming digital human,'' \url{https://github.com/lipku/livetalking}, 2024,
\newblock Accessed: 2025-03-16.

\bibitem{wav2lip}
K~R Prajwal and Rudrabha Mukhopadhyay,
\newblock ``{Wav2Lip}: A lip sync expert is all you need for speech to lip generation in the wild,'' \url{https://github.com/Rudrabha/Wav2Lip}, 2020,
\newblock Accessed: 2025-03-16.

\bibitem{zhang2024musetalk}
Yue Zhang, Minhao Liu, Zhaokang Chen, Bin Wu, Yubin Zeng, Chao Zhan, Yingjie He, Junxin Huang, and Wenjiang Zhou,
\newblock ``Musetalk: Real-time high quality lip synchronization with latent space inpainting,''
\newblock {\em arXiv preprint arXiv:2410.10122}, 2024.

\bibitem{metahumanstream}
tsman,
\newblock ``{metahuman-stream}: Real-time streaming digital human based on nerf,'' \url{https://github.com/tsman/metahuman-stream}, 2023,
\newblock Accessed: 2025-03-16.

\bibitem{rtmp}
{Adobe Systems Incorporated},
\newblock ``{Real-Time Messaging Protocol (RTMP) Specification},'' 2002,
\newblock Accessed: March 16, 2025.

\bibitem{webrtc}
{IETF and W3C},
\newblock ``{Web Real-Time Communication (WebRTC) Standard},'' 2011,
\newblock Accessed: March 16, 2025.

\bibitem{synthesia}
{Synthesia},
\newblock ``{Synthesia: AI Video Generation Platform},'' 2017,
\newblock Accessed: March 16, 2025.

\bibitem{vtubestudio}
Vincent Diener,
\newblock ``{VTube Studio: Live2D VTuber Streaming Software},'' 2021,
\newblock Accessed: March 16, 2025.

\bibitem{guo2024lightrag}
Zirui Guo, Lianghao Xia, Yanhua Yu, Tu~Ao, and Chao Huang,
\newblock ``Lightrag: Simple and fast retrieval-augmented generation,''
\newblock 2024.

\bibitem{gao2023funasr}
Zhifu Gao, Zerui Li, Jiaming Wang, Haoneng Luo, Xian Shi, Mengzhe Chen, Yabin Li, Lingyun Zuo, Zhihao Du, Zhangyu Xiao, et~al.,
\newblock ``Funasr: A fundamental end-to-end speech recognition toolkit,''
\newblock {\em arXiv preprint arXiv:2305.11013}, 2023.

\bibitem{edge-tts}
rany2,
\newblock ``{edge-tts}: Use microsoft edge's online text-to-speech service from python without needing microsoft edge or windows or an api key,'' \url{https://github.com/rany2/edge-tts}, 2024,
\newblock Accessed: 2025-03-16.

\end{thebibliography}

\end{document}